# Effects of thermal radiation heat transfer on flame acceleration and transition to detonation in particle-cloud flames

M.A. Liberman [1]  M.F Ivanov, A. D. Kiverin, [2]

[1] NORDITA, AlbaNova University Center, Roslagstullsbacken 23, SE-10691 Stockholm, Sweden
[2] Joint Institute for High Temperatures, Russia

**Abstract**

The current work examines regimes of the hydrogen-oxygen flame propagation and ignition of mixtures heated by radiation emitted from the flame. The gaseous phase is assumed to be transparent for the radiation, while the suspended particles of the dust cloud ahead of the flame absorb and reemit the radiation. The radiant heat absorbed by the particles is then lost by conduction to the surrounding unreacted gaseous phase so that the gas phase temperature lags that of the particles. The direct numerical simulations solve the full system of two phase gas dynamic time-dependent equations with a detailed chemical kinetics for a plane flames propagating through a dust cloud. It is shown that depending on the spatial distribution of the dispersed particles and on the value of radiation absorption length the consequence of the radiative preheating of the mixture ahead of the flame can be either the increase of the flame velocity for uniformly dispersed particles or ignition either new deflagration or detonation ahead of the original flame via the Zel'dovich gradient mechanism in the case of a layered particle-gas cloud deposits. In the latter case the ignited combustion regime depends on the radiation absorption length and correspondingly on the steepness of the formed temperature gradient in the preignition zone that can be treated independently of the primary flame. The impact of radiation heat transfer in a particle-laden flames is of paramount importance for better risk assessment and represents a route for understanding of dust explosion origin.

---

*) Corresponding author:  E-mail: mliber@nordita.org

# 1. Introduction

Hydrogen has emerged as an important fuel in a number of diverse industries as a means of achieving energy independence and to reduce emissions. Nowadays, when hydrogen technologies and fuel cells are penetrating the market in a number of applications, extensive research is still needed for effectively addressing the high-risk technological barriers in a pre-competitive environment. Wide spread deployment and use of hydrogen and hydrogen-based technologies can occur only if hydrogen safety issues have been addressed in order to ensure that hydrogen fuel presents at least the same level of hazards and associated risk as conventional fuel technologies. Hazard identification is the necessary step to ensure the full and safe utilization of hydrogen in either hydrogen safety engineering or risk assessment. The purpose of the hazard identification is to identify all events that can affect facility operation leading to a hazard to individuals or property. Since hydrogen is an extremely flammable and easily detonable gas when mixed with air over a wide range of composition, explosion hazards associated with the production, transportation and storage of hydrogen must be resolved to a sufficient confidence level and the key challenges facing the future widespread use of hydrogen are safety-related issues.

The hazardous potential of hydrogen-air and hydrogen-oxygen mixtures has been extensively studied and a huge number of experimental, theoretical and numerical studies inspired by their importance for industrial safety had been taken in attempt to understand nature of the explosion [1-9]. Although most accidental explosions are deflagrations, in the worst-case scenario, the flame acceleration can lead to deflagration-to-detonation transition (DDT). Depending on the mixture characteristics, such as concentrations, temperature, pressure and flow geometry, combustion process can undergo strong flame acceleration and deflagration-to-detonation transition (DDT). These regimes are characterized by high burning rates and consequently by high pressure loads. The resulting detonation is extremely destructive, can induce pressures up to or above 10 MPa, and therefore can have especially catastrophic consequences in a variety of industrial and energy producing settings related to hydrogen. Since the discovery of detonation more than 150 years ago, a huge number of experimental, theoretical and numerical studies had been taken in attempt to understand nature of the transition from deflagration to detonation (see e.g. [11, 12] and references within). These studies are inspired by their importance for industrial and nuclear power plants safety as well their potential application for micro-scale propulsion and power devices [13, 14]. Despite many years of substantial achievements in the area of the flame acceleration and DDT, still many specific aspects of the problem remain unclear. Dependence of the potential danger of the combustion process appeared to be very sensitive to the geometrical conditions of the processes, mostly to the confinement and to the congestion of the volume. Currently a unified physical model and corresponding numerical instrument which can be used over the entire range of phenomena is not available. Numerous combustion models are usually addressing only specific regime or phenomenon and are applicable only in their domain of validity.

The hazardous potential of hydrogen-oxygen and hydrogen-air mixtures has been extensively studied assuming a perfect mixture of hydrogen fuel and oxidant. Since the pioneering studies by Shchelkin, Zeldovich, Oppenheim and their co-authors [15-18] there has been a continuous efforts aiming to elucidate a reliable physical mechanism explaining DDT. From the beginning there was widely spread opinion that turbulence play a key role in the flame acceleration and DDT. A common belief was that a fast flame acceleration and the transition to detonation can occur only for strongly turbulent flames. Since the very first DDT studies it was known that the presence of obstacles increases the flame acceleration and shortens considerably the run-up distance. The experiments demonstrated that a flame accelerates more rapidly toward the open end of a duct if it passes through an array of turbulence-generating baffles. This presumably was the reason why the first attempts to explain DDT were associated with turbulent flames and were based on the assumption that DDT might occur only in the case of turbulent flames. Channels with rough walls or obstacles are often used to study DDT since it is believed that in this case the run-up distance is more or less fixed and controlled by turbulence [19, 20]. However, recent large scale experiments [21] with different kind of mixtures ($C_2H_4$-air, $CH_4$-air, $C_3H_8$-air, $H_2$-air) suggest that the self-acceleration mechanism of the flame may be much better represented by flame instabilities than by turbulence build-up. All the same DDT is easily observed in channels with smooth walls [22-24] and in thin capillary tubes [25]. The first explanation of the flame acceleration in tubes with no-slip walls before the DDT occurred was given by Zel'dovich [26]. In his detailed analysis of the Shchelkin's experimental results Zel'dovich has pointed that turbulence is not a primary factor responsible for flame acceleration in a smooth-walled channel and sequential detonation formation. Explaining the nature of the flame acceleration in the DDT events Zel'dovich emphasized that the flame acceleration in a tube with no-slip walls is due to stretching of the flame front caused by the flame interaction with a nonuniform velocity field of the upstream flow, while turbulence plays a supplementary role if any depending on the experimental conditions. Although the qualitative picture of the DDT is more or less clear, however a quantitative theory and the physical mechanism of DDT are still poorly understood and requires better theoretical and physical interpretation. With the advance in scientific computing, research has been shifted towards the use of computational approaches. Nowadays, numerical simulations can provide a qualitative picture of the basic processes from ignition and flame acceleration up to the transition to detonation. The reviews [11, 12] summarize the numerical efforts undertaken in the past decades to understand the deflagration-to-detonation mechanism in a highly reactive gaseous mixtures (e.g. hydrogen/air, acetylene/air) using a one-step Arrhenius chemical model. The conclusion drawn from these studies was that the mechanism of DDT is the Zeldovich gradient mechanism [27] involving gradient of reactivity. However, as an unsteady process, DDT involves multiple processes of vastly different scales. Among them, complex chemical reactions play a first-order controlling role for gaining scientific insight into the mechanism of DDT [28]. The numerical study of the Zeldovich gradient mechanism using a detailed chemical reaction models for hydrogen/oxygen and hydrogen/air [29, 30] has shown that the minimum scale of the temperature gradient (the length-scale of the temperature inhomogeneity) capable to initiate detonation exceeds size of the hot spots formed in the unreacted material ahead of accelerating flame by orders of magnitude. Recent 2D and 3D numerical simulations of the flame acceleration and DDT in hydrogen-oxygen mixtures that have taken into account a detailed chemical kinetics have revealed an adequate mechanism of DDT [24, 31-34].

An important problem of "hydrogen safety" is connected with leakage of the hydrogen and its further explosion. Hydrogen release might occur on storage, transport and handling. Once a flammable mixture forms, it can be ignited by a variety of uncontrolled means. An ignition is likely for all forms of releases due to the wide range of flammability and low ignition energy. The destructive power of the resulting explosion depends on the volume of the reactive mixture, its composition, and the geometries of confinement. In most practical cases ignition arises from a small area of combustible mixture, as an accidental ignition e.g. from an electric spark, or any local heating and starts as a laminar flame. The flame evolution can result in substantial flame acceleration, depending on the geometries of confinement, e.g. friction of the wall results in the flame front stretching, obstacles may result in flame turbulization, etc. Mechanisms of ignition by transient energy deposition and different scenarios of a detonation initiation were investigated in [35].

Most of studies on dust explosions have been performed to examine the characteristics or the indices of dust explosions in a closed vessel [9]. The fundamental mechanisms of flame propagation in dust suspension, however, have not been sufficiently studied and the limited experimental results known in the literature are often contradict each other. In particular, the influence of radiative heat transfer on the rate of flame propagation and duct cloud explosions is not yet fully understood. Combustion of the hydrogen-oxygen and hydrogen-air mixtures was studied largely at standard environmental conditions, however new technologies often dictate substantial variation of the conditions which have to be taken into account for the safety analysis. Among others, the fundamental properties of hydrogen mixtures with dusts of fine particles have to be considered. Non-uniformities of the gas distribution and small, micron-size solid particles suspended in the gas mixture can considerably affect regime of combustion in some cases leading to the strong flame acceleration and DDT. However, only limited amount of the experimental data are available on the behavior of the hydrogen in the presence of suspended particles related to the problem of dust explosion. Majority of the previous studies used a one-step chemical reaction model and were mainly focused on the deflagration-to-detonation transition (DDT) in a gaseous mixtures in attempt to understand mechanism of the detonation formation [36]. More than 50 years ago Essenhigh and Csaba [37] have indicated that radiation heat transfer plays a very significant role in the behavior of a dust cloud's planar flame. In earlier studies the flame dynamics affected by the radiative preheating has been investigated using asymptotic methods and a one-step Arrhenius chemical model with high activation energy [38-40]. Majority of research has been focused coal combustion with two aspects of practical interest: the production of volatiles due to thermal decomposition of coal dust and the ignition conditions [41-45]. The combustible volatiles can react and release energy, which in turn may contribute to the heat-up of the particles, enhance the combustion energy release due to energy feedback mechanism resulting in an explosion [45, 46]. For the coal-dust suspension air filling the coal-fired burners and for rocket engines using the solid or fluid fuels as well as for coal-fire mining safety both the ignition and combustion evolution are of paramount importance. Although there is an obvious lack of radiation measurements in large scale premixed systems related to explosions and the actual level of thermal radiation emitted from the flame remains conjectural, recent experiments have shown that the dust cloud flame propagation is strongly influenced by the thermal radiation [47, 48]. The data presented permits an assessment of the plausibility of combustion initiation due to forward thermal radiation. Although some models were developed recently, they were only for specific dusts (mainly for a coal dust flames) and might not be applicable generally [43, 48-50]. Identities of the important flame process particularly the controlling regimes of the flame propagation process, the knowledge of the role of the radiation heat transfer is highly demanded due to the raised likelihood to meet such conditions in the accidental conditions.

In the present paper we consider the influence of thermal radiation emanating from the hot combustion products on the rate of flame propagating in hydrogen gaseous mixture suspended with fine particles. Numerical and experimental studies show that hydrogen accumulation leads either to a stratified distribution of concentration or to the formation of a homogeneous layer if the convective flows at the top of the enclosure are high enough. Thus, we consider two possible scenarios: homogenous and a stratified (layers) nonuniform distribution of the suspended particles. In the case of non uniform dispersion of the particles, which is typical for e.g. dust layers [7, 9, 43-45], the time of the radiative heating is longer and the radiative preheating can be sufficient to ignite the surrounding combustible mixture ahead of the flame via the Zel'dovich gradient mechanism. In the latter case the ignited regime of combustion can be either deflagration or detonation depending on the radiation absorption length, which in turn determines steepness of the temperature gradient in the non uniform temperature distribution formed within the preheat zone. The effects of thermal radiation heat transfer on flame acceleration and transition to detonation in dusty-laden hydrogen mixtures analyzed theoretically and using high resolution numerical simulations for combustible materials whose chemistry is governed by a detailed chemical reaction model for chain-branching reactions.

**2. Problem setup. Physical and numerical model**

We consider a plane flame propagating from the closed to the open end of the duct in a particle-cloud hydrogen-oxygen mixture. For a pure gaseous flames the radiation emanated by high temperature combustion products is not important because the radiation absorption length in a gaseous mixture at normal conditions is very large, so that the gas is usually considered as fully transparent for the radiation and the radiation heat transfer does not influence the flame dynamics. For example, at normal pressure P=1atm the cross sections of the Thompson scattering and "bremsstrahlung" processes in the air are very small, $(10^{-24} \div 10^{-25})cm^2$, so that the mean free pass of a photon is more than tens meters. The radiation heat losses of the high temperature combustion products are small compared with the gaseous thermal conduction heat transfer to the tube walls and can be neglected. In the traditional theoretical studies of combustion the heat is transferred by the conduction and/or convection while the radiation heat transfer is negligible because the energy transferred through radiation is generally too small to affect the velocity of combustion wave. The situation changes drastically if the gaseous mixture is seeded by the particles. Particles are absorbed the radiant heat flux emanated from the flame, their temperature increased and then they lost heat by conduction to the surrounding unreacted gaseous phase, so that the gas phase temperature lags that of the particles. The radiation preheating of the gas ahead of the flame affects the flame dynamics resulting in the increase of the flame velocity or in the ignition of either deflagration or detonation in the "distant" particle seeded layer ahead of the flame in the case of a nonuniformly dispersed particle-cloud.

*2.1. Governing equations*

The governing equations for the gaseous phase are the one-dimensional, time-dependent, multispecies reactive Navier-Stokes equations including the effects of compressibility, molecular diffusion, thermal conduction, viscosity, gas-particles momentum and energy exchange and detailed chemical kinetics for eight reactive species $H_2$, $O_2$, H, O, OH, $H_2O$, $H_2O_2$, and $HO_2$ with subsequent chain branching, production of radicals, energy release and heat transfer between the particles and the gas phase. The governing equations for the gaseous phase are

$$\frac{\partial \rho}{\partial t} + \frac{\partial (\rho u)}{\partial x} = 0, \tag{1}$$

$$\frac{\partial Y_i}{\partial t} + u \frac{\partial Y_i}{\partial x} = \frac{1}{\rho} \frac{\partial}{\partial x}\left( \rho D_i \frac{\partial Y_i}{\partial x} \right) + \left( \frac{\partial Y_i}{\partial t} \right)_{ch}, \tag{2}$$

$$\rho \left( \frac{\partial u}{\partial t} + u \frac{\partial u}{\partial x} \right) = -\frac{\partial P}{\partial x} + \frac{\partial \sigma_{xx}}{\partial x} - \rho_p \frac{(u - u_p)}{\tau_{St}}, \tag{3}$$

$$\rho \left( \frac{\partial E}{\partial t} + u \frac{\partial E}{\partial x} \right) = -\frac{\partial (Pu)}{\partial x} + \frac{\partial}{\partial x}(\sigma_{xx} u) + \frac{\partial}{\partial x}\left( \kappa(T) \frac{\partial T}{\partial x} \right) +$$

$$+ \sum_k h_k \left( \frac{\partial}{\partial x}\left( \rho D_k(T) \frac{\partial Y_k}{\partial x} \right) \right) + \rho \sum_k h_k \left( \frac{\partial Y_k}{\partial t} \right)_{ch} - \rho_p u_p \frac{(u - u_p)}{\tau_{St}} - \rho_p c_{P,p} Q, \tag{4}$$

$$P = R_B T n = \left( \sum_i \frac{R_B}{m_i} Y_i \right) \rho T = \rho T \sum_i R_i Y_i, \tag{5}$$

$$\varepsilon = c_v T + \sum_k \frac{h_k \rho_k}{\rho} = c_v T + \sum_k h_k Y_k, \tag{6}$$

$$\sigma_{xx} = \frac{4}{3} \mu \left( \frac{\partial u}{\partial x} \right), \tag{7}$$

where we use standard notations: P, $\rho$, u represent pressure, mass density and flow velocity of the gas phase, $Y_i = \rho_i / \rho$ is the mass fractions of the species, $E = \varepsilon + u^2/2$ is the total energy density, $\varepsilon$ is the internal energy density, $R_B$ is the universal gas constant, $m_i$ is the molar mass of i-species, $R_i = R_B / m_i$, n is the gas molar density, $\sigma_{ij}$ is the viscous stress tensor, $c_v = \sum_i c_{vi} Y_i$ is the constant volume specific heat, $c_{vi}$ is the constant volume specific heat of i-species, $h_i$ is the enthalpy of formation of i-species, $\kappa(T)$ and $\mu(T)$ are the coefficients of thermal conductivity and viscosity, $D_i(T)$ is the diffusion coefficients of i-species, $(\partial Y_i / \partial t)_{ch}$ is the variation of i-species concentration (mass fraction) in chemical reactions, $\rho_p = m_p N_p$ is mass density of the suspended particles, $N_p$ is the particles number density, $u_p$, $r_p$, $m_p$ are velocity, radius and mass of the spherical particle, $\tau_{St} = m_p / 6\pi\mu r_p$ is the Stokes time, Q is thermal energy exchange between the gas phase and the particles, $c_{P,p}$ is the constant pressure specific heat of the particle material.

The changes in concentrations of the mixture components due in the chemical reactions are defined by the solution of equations of chemical kinetics

$$\frac{dY_i}{dt} = F_i(Y_1, Y_2, ... Y_N, T), \quad i = 1, 2, ... N. \tag{8}$$

The right hand parts of Eq. (8) contain the rates of chemical reactions for the reactive species. The reaction mechanism is the standard reduced chemical kinetic scheme for hydrogen/oxygen consisting of 19 elementary reactions of the Arrhenius type with pre-exponential constants and activation energies presented in [51]. This reaction scheme for a stoichiometric $H_2$-$O_2$ mixture has been tested in many applications and to a large extent adequate to complete chemical kinetics well describing the main features of the $H_2$-$O_2$ combustion. The computed thermodynamic, chemical, and material parameters using this chemical scheme are in a good agreement with the flame and detonation parameters measured experimentally. For $P_0 = 1.0 \, \text{bar}$ we obtained for the laminar flame velocity, the flame thickness, adiabatic flame temperature and the expansion coefficient (the ratio of densities of the unburned gas, $\rho_u$, and the combustion products, $\rho_b$): $U_f \approx 12 \, \text{m/s}$, $L_f = 0.24 \, \text{mm}$, $T_b = 3012K$, $\Theta = \rho_u / \rho_b = 8.36$, correspondingly, and for temperature and velocity of CJ-detonation $T_{CJ} = 3590K$, $U_{CJ} = 2815 m/s$.

The transport coefficients were calculated using the gas kinetics theory [52]. The gaseous mixture viscosity coefficients are

$$\mu = \frac{1}{2}\left[\sum_i \alpha_i \mu_i + \left(\sum_i \frac{\alpha_i}{\mu_i}\right)^{-1}\right], \tag{9}$$

where $\alpha_i = \frac{n_i}{n}$ is the molar fraction, $\mu_i = \frac{5}{16}\frac{\sqrt{\pi \hat{m}_i kT}}{\pi \Sigma_i^2 \tilde{\Omega}_i^{(2,2)}}$ is the viscosity coefficient of $i$-species, $\tilde{\Omega}^{(2,2)}$ - is the collision integral which is calculated using the Lennard-Jones potential [52], $\hat{m}_i$ is the molecule mass of the i-th species of the mixture, $\Sigma_i$ is the effective molecule size. The thermal conductivity coefficient of the gas mixture is

$$\kappa = \frac{1}{2}\left[\sum_i \alpha_i \kappa_i + \left(\sum_i \frac{\alpha_i}{\kappa_i}\right)^{-1}\right]. \tag{10}$$

Coefficient of the heat conduction of i-th species $\kappa_i = \mu_i c_{pi} / \Pr$ can be expressed via the kinematic viscosity $\mu_i$ and the Prandtl number, which is taken $\Pr \approx 0.71 \div 0.75$.

The binary coefficients of diffusion are

$$D_{ij} = \frac{3}{8}\frac{\sqrt{2\pi kT \hat{m}_i \hat{m}_j / (\hat{m}_i + \hat{m}_j)}}{\pi \Sigma_{ij}^2 \tilde{\Omega}_{ij}^{(1,1)}(T_{ij}^*)} \cdot \frac{1}{\rho}, \tag{11}$$

where $\Sigma_{ij} = 0,5(\Sigma_i + \Sigma_j)$, $T_{ij}^* = kT/\varepsilon_{ij}^*$, $\varepsilon_{ij}^* = \sqrt{\varepsilon_i^* \varepsilon_j^*}$; $\varepsilon^*$ are the constants in the expression of the Lennard-Jones potential, and $\tilde{\Omega}_{ij}^{(1,1)}$ is the collision integral similar to $\tilde{\Omega}^{(2,2)}$ [52]. The diffusion coefficient of i-th species is

$$D_i = (1-Y_i)/\sum_{i \neq j} \alpha_i / D_{ij}. \tag{12}$$

A detailed description of the transport coefficients used for the gaseous phase and calculation of diffusion coefficients for intermediates is described in detail in [24, 31-33].

The equations of state of the fresh mixture and combustion products are taken with the temperature dependence of the specific heats, heat capacities and enthalpies of each species borrowed from the JANAF tables and interpolated by the fifth-order polynomials [53].

The solid particles are modeled using continuous hydrodynamic approximation. The particle-particle interactions is negligible for a small volumetric concentration of the particles and only the Stokes force between the particle and the gaseous phase must be taken into account. Thus, the equations for the suspended particles are:

$$\frac{\partial N_p}{\partial t} + \frac{\partial (N_p u_p)}{\partial x} = 0 \qquad (13)$$

$$\left(\frac{\partial u_p}{\partial t} + u_p \frac{\partial u_p}{\partial x}\right) = \frac{(u - u_p)}{\tau_{St}}, \qquad (14)$$

$$\left(\frac{\partial T_p}{\partial t} + u_p \frac{\partial T_p}{\partial x}\right) = Q - \frac{2\pi r_p^2 N_p}{c_{P,p} \rho_{p0}} \left(4\sigma T_p^4 - q_{rad}\right), \qquad (15)$$

where $T_p$ – temperature of the particles, $2\pi r_p^2 N_p \left(4\sigma T_p^4 - q_{rad}\right)$ - is the thermal radiation heat flux absorbed and reemitted by the particles. The heat transferred from the particle surface to surrounding gaseous mixture is

$$Q = (T_p - T) / \tau_{pg}, \qquad (16)$$

where $\tau_{pg} = 2r_p^2 c_{P,p} \rho_{p0} / 3\kappa Nu$ is the characteristic times of the energy transfer from the particle surface to surrounding gaseous mixture, $c_{P,p}$ and $\rho_{p0}$ are specific heat and the density of the particle material, $Nu$ is the Nusselt number [54].

For a one-dimensional plane problem the equation for the thermal radiation heat transfer in the diffusion approximation is [55, 56]:

$$\frac{d}{dx}\left(\frac{1}{\chi}\frac{dq_{rad}}{dx}\right) = -3\chi\left(4\sigma T_p^4 - q_{rad}\right), \tag{17}$$

where the radiation absorption coefficient is $\chi = 1/L = \pi r_p^2 N_p$, and $L = 1/\pi r_p^2 N_p$ is the radiation absorption length.

It should be noticed that although the emissivity of the gaseous combustion products is relatively low, however even small concentration of particles increases the emissivity considerably. From theoretical point of view the particle-laden mixture is optically-thick if the radiation absorption length is small compared with the domain size. Radiation emanates from both gaseous combustion products and suspended particles, which are at high temperatures and hence high radiation levels are expected. The thermal radiation flux emanating from both high temperature combustion products with seeded particles depends on the emissivity of the burned volume, which relates to the concentration of particles in combustion products and was assessed to be close to theoretical values of blackbody radiation for a burned gas temperature [47]. Thus, we assume the blackbody radiative flux emitted from the flame front, $q_{rad}(x = X_f) \approx \sigma T_b^4$, where $\sigma = 5.6703 \; 10^{-8} \, W/m^2 K^4$ is the Stefan-Boltzmann constant, and $T_b \approx 3000K$ is temperature of the hydrogen-oxygen combustion products. The calculations were carried out for stoichiometric hydrogen-oxygen at initial pressure $P_0 = 1 atm$ with an identical micron-scale inert spherical particles suspended in the gaseous mixture. The mass loading parameter was taken small, $\zeta = \rho_p/\rho \ll 1$, so that there is only one way of a momentum coupling of the particles and the gaseous phase. Therefore, only heating of the particles by the radiation and the heat transfer from the particles to the gas phase ahead of the flame will influence the flame dynamics. This choice of the parameters allows us to distinguish the effect of the radiation preheating on the flame dynamics.

*2.3. Numerical algorithms*

High resolution numerical simulations were performed to study the effect of thermal radiation heat transfer and the interaction of a stoichiometric hydrogen-oxygen and hydrogen-air flames with a dust cloud composed of either uniformly dispersed or stratified (layered) particles. An exhaustive description of the convergence and resolution tests of the numerical method have been performed in [24, 31-33]. The governing equations are solved using the method developed originally by Gentry, Martin and Daly [57] and afterwards modified and implemented for various hydrodynamic problems by Belotserkovsky and Davydov [58]. The method is based on splitting of the Eulerian and Lagrangian stages and known as Coarse Particle Method (CPM). The algorithm was further modified by Liberman et al. [59] so that a high numerical stability of the method is achieved if the hydrodynamic variables are transferred across the grid boundary with the velocity, which is an average value of the velocities in neighboring grids. The second order in space solver was thoroughly tested and successfully used for modeling knock appearance in SI-engines [59, 60], to study the flame acceleration in tubes with non-slip walls and DDT [31-33] and various problems of transient combustion, e.g. ignition of different combustion modes [29, 30, 35]. The system of chemical kinetics equations represents a stiff system of differential equations and was solved using standard Gear's method [61]. The developed algorithm was implemented using the FORTRAN-90. An additional convergence and resolution tests were carried out to verify that the observed phenomena of the radiation preheating and impact of the radiation heat transfer in particle-laden flame were correctly caught remaining unchanged for increased resolution. The main specific feature of the reacting flows is the presence of the reaction fronts that must be resolved fine not only to diminish the scheme viscosity influence but to resolve their structures and interactions with the compression and shock waves that can arise in the compressible flow. The fine grids according to the physical limitations were used. To resolve structure of the hydrogen-oxygen flame front which thickness is about 0.24mm at normal conditions the grid cells less than 0.01mm were used to avoid unphysical couplings with the shocks that are usually smoothed over 5-6 cells even by the high-order schemes with limiters. The convergence and resolution tests were carried out to verify that finer grid resolutions, up to 0.005mm provides quantitative reliability of the method for calculating the flames propagating at elevated temperatures and pressures and interacting with the compression and shock waves that can arise in the compressible flow. In particular it was demonstrated that the computational method reproduces with high accuracy analytical solution for discontinuity decay in air for different driver pressures and a so-called "Blast wave" test by Woodward and Colella [62].

### 3. Effect of radiation preheating for uniformly dispersed particles

3.1. *The flame acceleration due to radiation preheating*

We consider the hydrogen-oxygen flame propagating in the mixture with uniformly dispersed identical solid spherical particles of radius $r_p = 0.75 \mu m$, and the particle mass density $\rho_{p,0} = 1 g/cm^3$. The thermal radiation emitted from the flame is absorbed and reemitted by the particles ahead of the flame and the intensity of the radiant flux decreases exponentially on the scale of the order of the radiation absorption length $L = 1/\pi r_p^2 N_p$. The time evolution of the temperature at the distance just 2mm ahead the flame front shown in Fig.1 was computed for different radiation absorption lengths: L=1, 2, and 4cm corresponding to the concentration of particles: $N_p = 5.7 \cdot 10^7; 2.85 \cdot 10^7; 1.4 \cdot 10^7 cm^{-3}$. The maximum temperature of the particles ahead of the flame and the maximum increase of the flame velocity achieved during the characteristic hydrodynamic time of the order $t \approx L/U_{f0}$, where $U_{f0}$ is the normal laminar flame velocity in pure gas mixture. It seen from Fig.1 that the maximum temperature of the radiation preheating depends weakly on the radiation absorption length. Although the local radiant heat flux absorbed by the particle is larger for smaller absorption length, but in the case of a larger absorption length particles absorb the radiant heat flux over a longer time until their arrival to the flame front. This difference for smaller and larger absorption lengths compensates the lesser local heating for a larger absorption lengths.

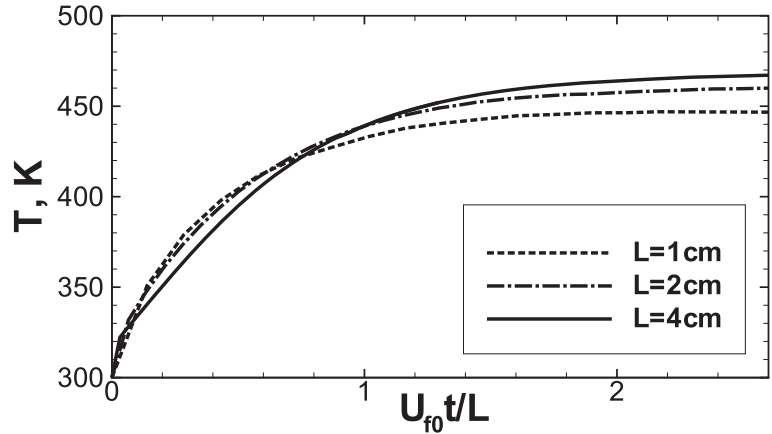

**Figure 1.** Time evolution of the gaseous temperature during radiative preheating at the distance 2mm ahead the flame front for different radiation absorption lengths, L=1, 2, and 4cm. On the x-axis time is in units $L/U_{f0}$.

Fig. 2 shows the profiles of the gaseous phase temperature which are established after the stationary flow was settled for the radiation absorption lengths, L=1, 2, and 4cm.

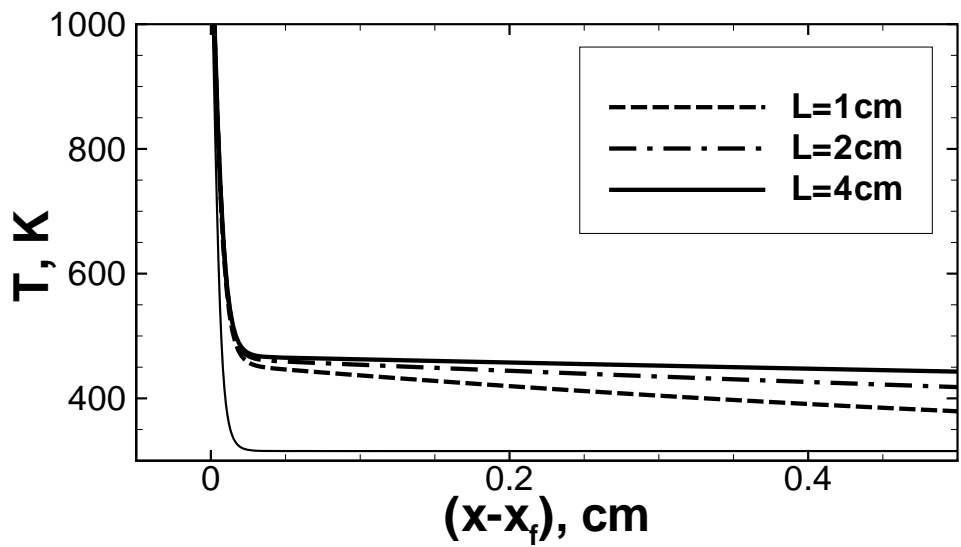

**Figure 2.** Temperature distribution ahead of the flame front calculated for uniformly distributed suspended particles and for different radiation absorption lengths, L=1, 2, and 4 cm. Thin line corresponds to the temperature profile for the laminar flame in a pure stoichiometric hydrogen/oxygen mixture.

The flame velocity increase with respect to the unburned gas due to the radiation preheating (Fig.2) is shown in Fig. 3. Notice that for a flame propagating from the closed end of a duct, the unburned gas ahead of the flame moves to the open end with the velocity $u = (\Theta - 1)U_f$, where $\Theta = \rho_u / \rho_b$ is the density ratio of the unburned $\rho_u$ and burned $\rho_b$ gas [63, 64]. The velocity of the flame with respect to the tube walls is $U_{fL} = \Theta U_f$ and with respect to the unburned gas the flame velocity is $U_f$.

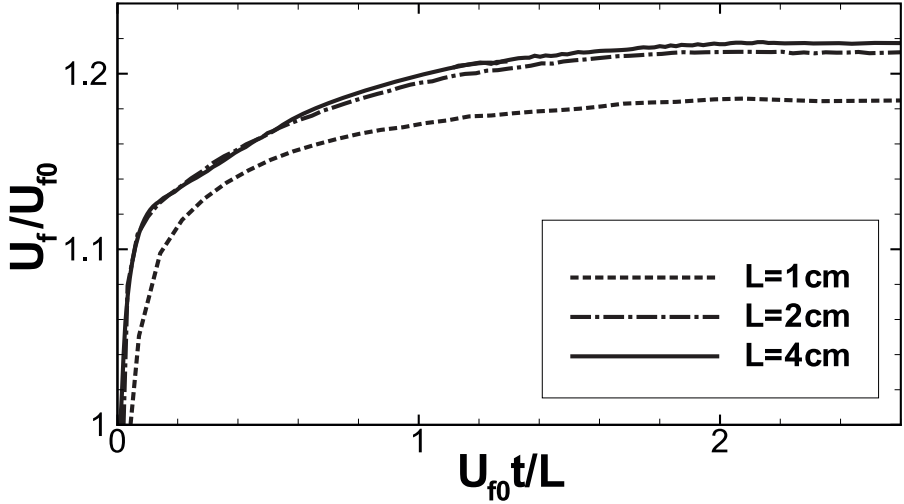

**Figure 3.** Computed time evolution of the flame velocity for the uniformly suspended micro-particles for different thermal radiation absorption lengths L. The flame velocity is normalized on the normal laminar flame velocity $U_{f0}$ in pure gas mixture. Time is in units $L/U_{f0}$.

The radiant heat absorbed by the particles is then lost by conduction to the surrounding unreacted gaseous phase and the gas phase temperature lags that of the particles. The characteristic time scales of the problem. For the parameters used in simulations the characteristic time scales are: $\tau_{St} = 2r_p^2 \rho_{p0} / 9\nu_g \rho_g \approx 10\mu s$, $\tau_{pg} = 2r_p^2 c_{P,p} \rho_{p0} / 3\kappa Nu \approx 0.3\mu s$, $L_f / U_f \approx 20\mu s$ and $L/U_f \approx 1ms$. Since the characteristic time of energy transfer between the particles and the gaseous phase $\tau_{pg}$ is much smaller than the gas-dynamic time scales $L/U_f$, the temperatures of the particles and the gaseous phase are approximately equal, $T_p \approx T$. The value of mass loading parameters are: $\varsigma = \rho_p / \rho = 0.2$; 0.1 and 0.05 for $L = 1, 2$ and 4cm, correspondingly. Because $\varsigma = \rho_p / \rho \ll 1$, the momentum coupling of the particles and the gaseous phase is small. Thus, only the radiation heating of particles and corresponding preheating of the gas mixture ahead of the flame influence the flame dynamics. The presence of inert particles is similar to dilution of combustible mixture with inert gas, decreasing the adiabatic temperature behind the flame. This effect is small for $\varsigma(c_{p,p}/c_{V,g}) \ll 1$. This choice of the parameters allows us to distinguish the effect of the radiation preheating on the flame dynamics.

Taken into account that the stationary flow is established during the time of the order $L/U_f$, it is straightforward to obtain an estimate for the maximum temperature increase of the gas ahead of the flame. In the coordinate system co-moving with the flame front the unburned mixture with suspended particles flows toward the flame with the laminar flame velocity $U_f$. The thermal radiation is appreciably absorbed by the particles, which are located at $x \leq L$ ahead of the flame front, and their temperature is defined by the energy equation

$$\rho_p c_{p,p} \frac{dT_p}{dt} = \sigma T_b^4 \exp\left(-\frac{x - U_f t}{L}\right) \pi r_p^2 N_p - \frac{\rho_p c_{p,p}}{\tau_{pg}}(T_p - T). \tag{18}$$

The gaseous phase temperature ahead of the flame increases due to the heat transferred from the particles to the surrounding gas

$$\frac{dT}{dt} = -\frac{(T - T_p)}{\tau_{gp}} \tag{19}$$

Taken into account that $T \approx T_p$ for $\tau_{pg} \ll L/U_f$ we obtain from (18) and (19)

$$\rho_p c_{p,p}(1 + 1/\xi)\frac{dT}{dt} = \sigma T_b^4 \frac{1}{L} \exp\left(-\frac{x - U_f t}{L}\right), \tag{20}$$

where $\xi = \tau_{gp}/\tau_{pg} = c_{V,g}/\varsigma c_{P,p}$.

The characteristic time of the radiation heating is approximately time of the Lagrangian particle arrival to the flame front, $t \approx L/U_f$. Then, the maximum temperature increase achieved in the mixture close to the flame front can be estimated as:

$$\Delta T = \sigma T_b^4 \frac{1}{U_f} \frac{(1-e^{-1})}{(\rho_p c_p + \rho c_{V,g})} \approx 0.63 \frac{\sigma T_b^4}{U_f(\rho_p c_p + \rho c_{V,g})}. \tag{21}$$

It can be readily observed from (21) that the maximum temperature increase due to the radiation preheating does not depend on the radiation absorption length (see also Fig.1). The maximum temperature increase given by (21) is in a good agreement with computed values, taking into account that the adiabatic flame temperature is smaller due to gas dilution by the inert solid particles, and that the photons emitted from the flame front are produced within the radiative layer near the flame front of the finite thickness.

3.2. *Structure of the flame; radiation dominated regime of the flame propagation*

It follows from Eq. (21) that the effect of the radiation preheating is enhanced for a flame with lower normal laminar flame velocity or for a lower initial gas density. The computed increase of the gas temperature and the corresponding increase of the flame velocity of the $H_2$-$O_2$ flames propagating at the initial pressures 1.0atm, 0.3atm and 0.2.atm are shown in Fig.4(a, b).

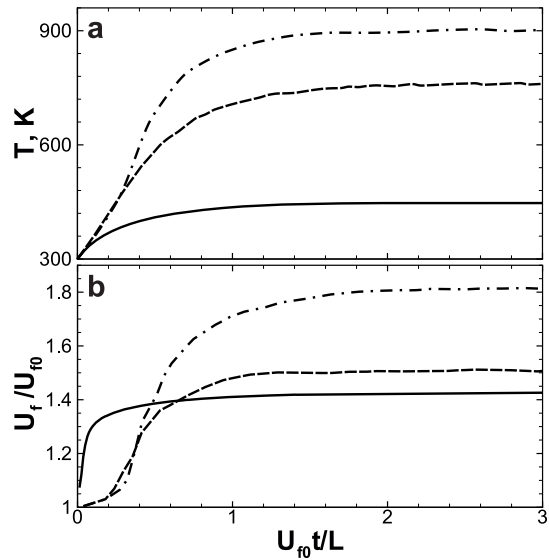

**Figure 4 (a, b).** Radiation preheating at the distance 2mm ahead the flame front (a) and the flame velocity increase (b) for different initial pressures of the for $H_2/O_2$ mixture: solid line - $P_0 = 1$atm, dashed line - $P_0 = 0.3$atm, dashed-doted line - $P_0 = 0.2$atm. The radiation absorption length is L=1cm. On the x-axis time is in units $L/U_{f0}$.

Fig. 5(a, b) shows the time evolution of the maximum temperature and the corresponding increase of the hydrogen-air flame velocity in comparison with these values for hydrogen-oxygen flame for initial pressures 1atm and radiation absorption length $L = 1 cm$. It is seen that for slower $H_2$-air flame the preheating is more effective and the flame velocity increased much more than for the $H_2$-$O_2$ flame.

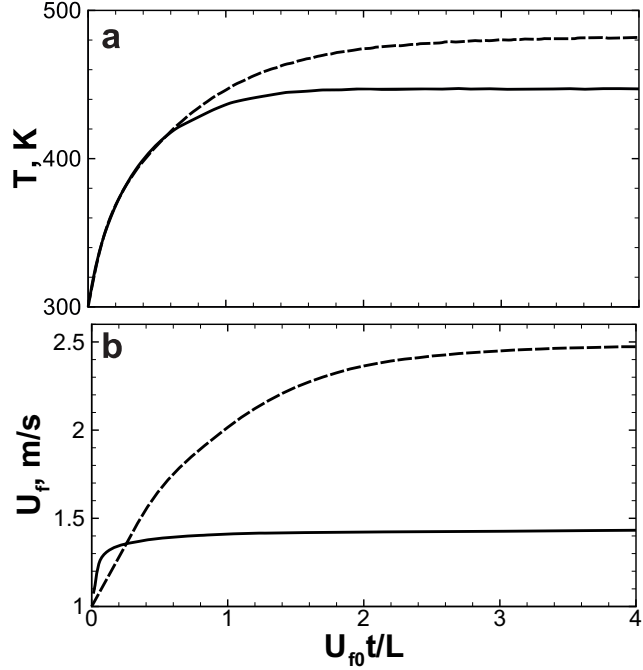

**Figure 5. (a):** Time evolution of the gaseous temperature during radiative preheating at the distance 2mm ahead the flame front for $H_2/O_2$ (solid line) and $H_2$/air (dashed line) flames for radiation absorption lengths, L=1cm, $P_0 = 1 atm$. On the x-axis time is in units $L/U_{f0}$. 5(**b**): The flame velocity increase for the conditions of Fig.5(a) for $H_2/O_2$ (solid line) and $H_2$/air (dashed line) flames for radiation absorption lengths, L=1cm, $P_0 = 1 atm$. The flame velocities are normalized on the corresponding normal laminar flame velocity $U_{f0}$ in pure gas mixture. Time is in units $L/U_{f0}$.

It is interesting to note that in the framework of a one-step chemical model the width of a flame decreases with the increase of the flame speed. Indeed, the velocity and structure of a laminar flame in the classical combustion theory [63, 64] is defined in the approximation of the linear equation of thermal conduction, which solution in a one-dimensional case is

$$T(x,t) \propto \frac{1}{2\sqrt{\pi \chi_g t}} \exp(-x^2/4\chi_g t), \qquad (22)$$

where $\chi_g = \kappa_g / \rho_g c_{g,p} \approx \text{const}$ is the thermal diffusivity coefficient. It follows from Eq. (22) that during time t the heat propagates at the distance $x \propto \sqrt{4\chi_g t}$. Taking this into account, it is straightforward to obtain an order-of-magnitude estimates for the speed and width of the laminar flame, which propagates due to thermal conduction:

$$L_f \propto \sqrt{\chi_g \tau_R}, \quad U_f \propto \sqrt{\chi_g / \tau_R}. \tag{23}$$

These expressions are based on the consideration that the characteristic heat diffusion time scale is much longer than the characteristic time $\tau_R$ of the heat release in the reaction, otherwise any small disturbances of the combustion wave will diffuse away resulting in the flame extinguishing. Thus, we arrive to the conclusion that $L_f U_f \propto \chi_g$. On the contrary, the increase of the flame velocity caused by the radiation preheating is accompanied by an increase of the flame front width. Fig.6 shows computed width of the flame depending on the radiation absorption length. Large values of L correspond to pure gas mixture with the flame front width $L_f = 0.24mm$ for $H_2$-$O_2$ at $P = 1atm$.

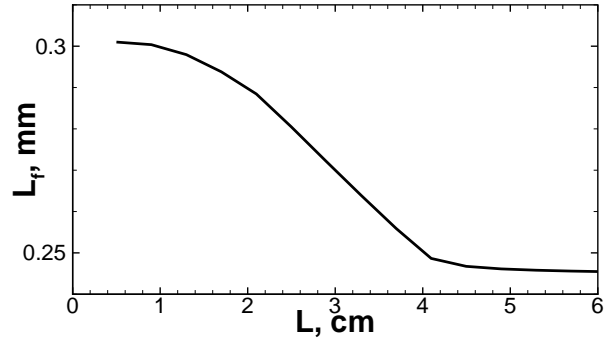

**Figure 6.** Width of the hydrogen-oxygen flame versus the radiation absorption length, $P_0 = 1atm$.

According to (21) for the flame with small enough velocity the maximum temperature of the unreacted gas ahead of the flame can exceed the ignition threshold. For a hydrogen-oxygen this is the crossover temperature, when the endothermic reaction stage passes to the fast exothermic stage. In this case radiation may dominate the conduction heat transfer and thermal radiation makes a decisive contribution to the overall energy transport. As an example Fig. 7(a, b) shows the computed time evolution of the maximum temperature increase ahead of the flame in methane-air with suspended particles and the corresponding flame velocity increase due to the radiation preheating for the radiation absorption length $L = 1 cm$. Because of much smaller velocity of the flame in methane-air, the radiation preheating time is much longer than for $H_2$-$O_2$ or $H_2$-air flames. In the calculations the reduced chemical reaction mechanism for methane-air combustion developed in [65] was used.

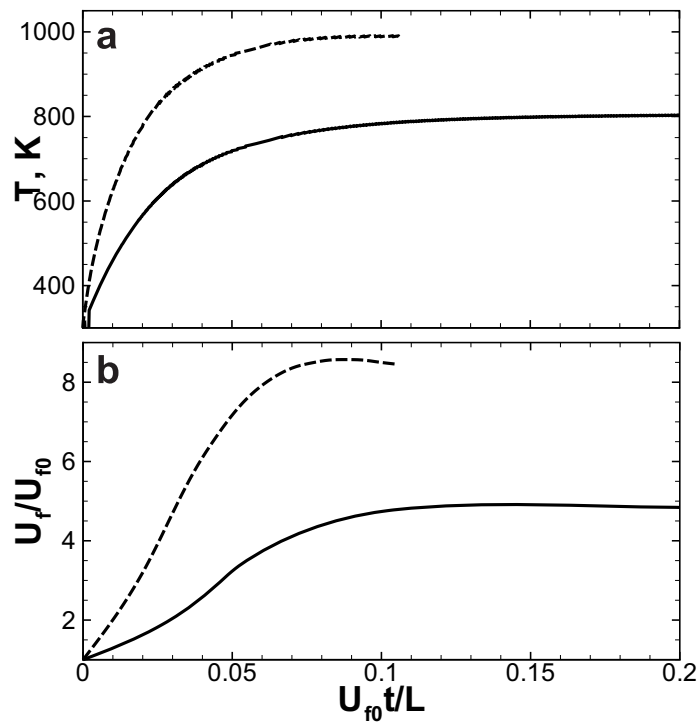

**Figure 7(a, b).** Radiation preheating at the distance 2mm ahead the methane-air flame (a) and the flame velocity increase (b) at initial pressure $P_0 = 1 atm$ (solid line) and $P_0 = 0.5 atm$ (dashed line). The radiation absorption length is L=1cm. On the x-axis time is in units $L/U_{f0}$.

The flame structure changes considerably if radiation dominates heat transfer process. In the framework of a one-step chemical model, the total enthalpy is constant inside of the flame which implies similarity of temperature and concentration of reagent distributions in the flame structure. This conclusion in the classical combustion theory obtained in the approximation of linear thermal conduction and when the coefficient of thermal diffusivity equals to the diffusion coefficient of reagents. On the contrary, if the radiation becomes dominating process, the heat propagation is defined by the nonlinear equation of heat conduction. In the diffusion approximation the coefficient of radiation thermal conductivity is

$$\kappa_{Rad} = 16\sigma T^3 L / 3 . \tag{24}$$

In this case the total enthalpy of the unburned mixture and combustion products remains equal but the total enthalpy is not constant inside of the flame and there is no similarity of the temperature and the density distributions. Solution of the thermal conduction equation for the nonlinear thermal conductivity (24) is much more gentle than that given by (22)

$$T(x) \propto T_0 \left(1 - x^2 / x_0^2\right)^{1/3} . \tag{25}$$

In this case the reaction can be ignited within the preheat zone, which size by the order of magnitude is the radiation absorption length L. Correspondingly, the flame thickness is by the order-of-magnitude equal to the radiation absorption length, i.e. $L_{fR} \propto L$, and the flame velocity can be estimated as $U_{f,rad} \propto (L / \tau_R) \approx (L / L_f) U_f \gg U_f$, where $L_f$ and $U_f$ are the width and velocity of the laminar flame given by Eq. (23). If radiation dominates the flame velocity exceeds significantly the laminar flame velocity in a pure gas mixture. As soon as combustion has been initiated by the primary particle-laden flame, the combustion generates secondary explosions in the mixture ahead of the flame. Such combustion wave will look like a sequence of thermal explosions and can be accompanied by a strong pressure increase. The necessary condition for the radiation to become dominating mechanism of a flame propagation is that the characteristic time of ignition in the mixture ahead of the flame to be small compare with the time the original flame passing through the radiation preheat zone, $\tau_{ign}(T_{cr}) \ll L / U_f$. It is unlikely that this condition is fulfilled for fast $H_2$-$O_2$ or $H_2$-air flames, but it is possible for a very slow flame, such as e.g. methane-air flame, where the process can be additionally enhanced due to volatile combustion of coal dust is involved. In order to have a notable effect ignition have to be formed well ahead of the advancing flame, thus relatively long time scale of the radiation preheating and length scales are essential. This is unlikely for a fast hydrogen-oxygen flame with uniformly dispersed particles, however it is possible for non-uniform distribution of suspended particles.

## 4. Ignition of deflagration and detonation by the radiation preheating for non-uniform concentration of particles

Non-uniform distribution of particles in a dust-air cloud may be formed when fine particles are raised either in the process of hydrogen leakage or by an expanding gas cloud or as result of a weak local explosion. In coal mines it is a well established phenomenon that the pressure wave of a weak methane explosion can disperse dust deposits leading to the formation of an non-uniformly suspended particles forming a layered dust-air cloud. We consider a non-uniform distribution of the particles concentration shown schematically in Fig.8.

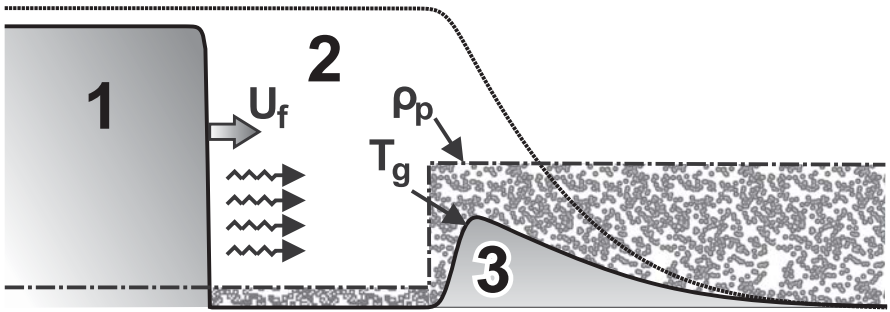

**Figure 8.** Scheme of the radiation preheating of the gaseous mixture inside the gas-particles cloud ahead the flame front. 1- high temperature combustion products; 2 - "gap" with lower concentration of particles; 3 –cloud of particles; $T_g$ -temperature of the radiative preheated gas.

Concentration of the particles immediately ahead of the flame (gap 2 in Fig. 8) is relatively low so that the radiation absorption length here is much larger then the gap width. Below we assume that the "gap" between the flame and the left boundary of the particles-cloud mixture is transparent for the radiation so that the radiant heat flux is fully absorbed in the layer 3. If time of the flame arrival to the boundary of the layer 3 is long enough, so that the temperature of the particles and the surrounding mixture can rise up to the ignition value before the flame arrival, which is about 1ms for $H_2/O_2$ flame, then the maximum temperature ($T_g$ in Fig. 8) within the temperature gradient established due to the radiative preheating exceeds the crossover value which is for hydrogen/oxygen at 1atm is 1050±50K. The corresponding width of the gap is in the range of a few centimeters. What kind of combustion regime is ignited via the Zel'dovich's gradient mechanism in the denser layer depends on the radiation absorption length and, correspondingly on the steepness of the formed temperature gradient.

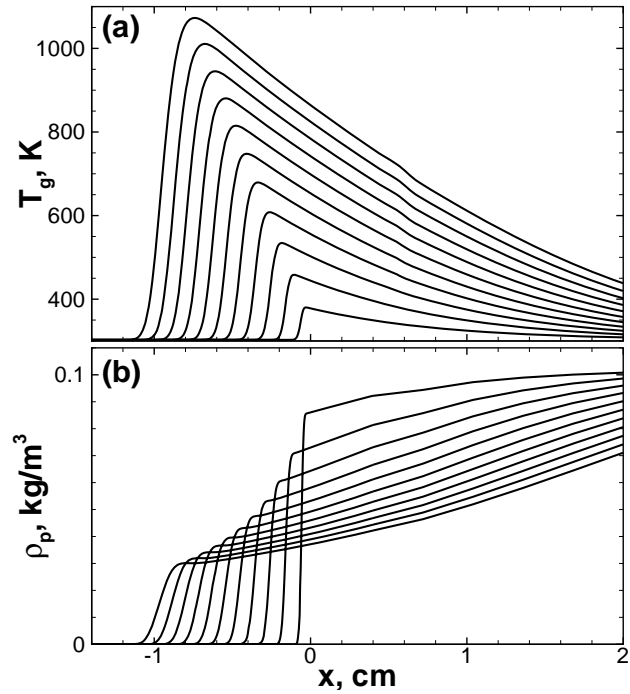

**Figure 9.** Temporal evolution of the gaseous temperature (a) and the mass density of the suspended solid particles (b) profiles during radiative preheating inside the gas-particles cloud ahead of the propagating flame. Profiles are shown with the time intervals of 50μs. For initial stepwise particles density profile, $N_p = 2.5 \cdot 10^7 \text{cm}^{-3}$, $r_p = 1\mu m$.

The temperature gradient, which is established due to the radiative preheating of the mixture in the gas-particle cloud, depends mainly on the radiation absorption length and it is also modified during the expansion of heated up mixture. Since the characteristic acoustic time in the preheat zone is much smaller then the time of the radiative heating up to crossover temperature, the pressure is equalized within the region heated by the radiation, and the temperature gradient is formed at the almost constant pressure $P \approx P_0 = 1$ atm. Classification of the combustion regimes in hydrogen-oxygen and hydrogen-air mixtures initiated by the initial temperature gradient via the Zel'dovich's gradient mechanism [27] has been investigated by Liberman et al. [29, 30] using a detailed chemical kinetic models.

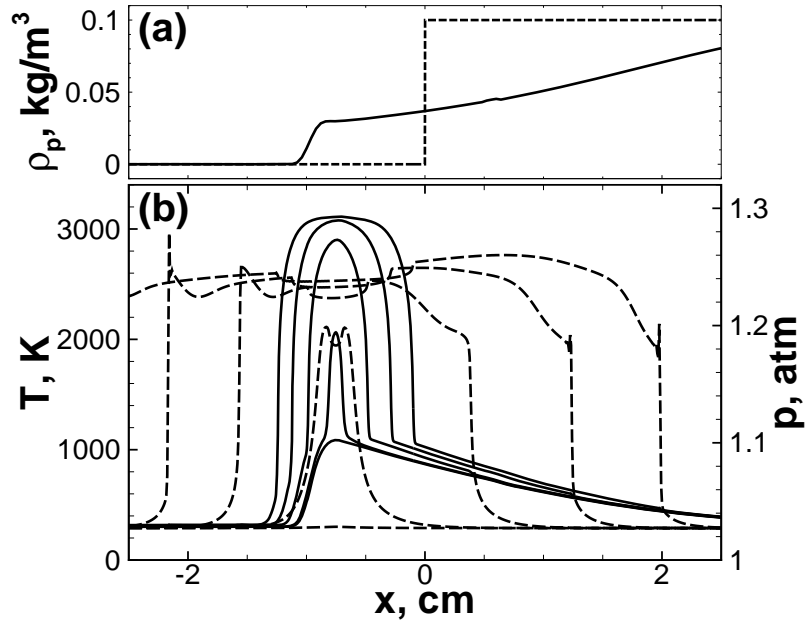

**Figure 10**. Temporal evolution of the gaseous temperature profiles (middle frame) and pressure profiles (bottom frame) during the slow combustion wave formation in the vicinity of the margin of the gas-particles cloud far ahead the propagating flame front, $t_0 = 900 \mu s$, $\Delta \tau = 50 \mu s$. The upper frame shows the distribution of particles mass density: the initial stepwise density profile (dashed line) and density profile at time instant $t_0$ prior to the ignition (solid line).

Figure 9(a, b) shows the computed time evolution of the gaseous temperature (9a) and the particles density profile (9b) during the preheating when the maximum temperature raised up to the crossover value and the final temperature gradient is formed at $t_0 \approx 900 \mu s$. The calculations were performed for the initial stepwise density of particles of radius $r_p = 1 \mu m$ with the maximum concentration of the particles within the gas-particles layer $N_p = 2.5 \cdot 10^7 \text{cm}^{-3}$. The scale of the temperature gradient in Fig. 9, which is formed near the left boundary of the particle-gas cloud is $\Delta = (T^* - T_0)/|dT/dx| \approx 1 \text{cm}$, which is close to the value of the radiation absorption length $L = 1/\pi r_p^2 N_p \approx 1.2 \text{cm}$. According to the classification of combustion regimes [29, 30] initiated by the initial temperature gradient in hydrogen oxygen at 1 atm, this temperature gradient can ignite a deflagration. The time evolution of the gaseous temperature profiles presented in the middle frame of Fig. 10 shows the development of the spontaneous reaction wave for the same conditions as in Fig. 9 and the formation of a deflagration wave. The dashed line in the upper frame of Fig. 10 shows the initial profile of the particle concentration while the solid line shows the particles number density profile formed at the instant $t_0 = 900 \mu s$ just prior to the ignition, when the temperature gradient with the maximum temperature $T^* = 1050 \text{ K}$ is formed (the first temperature profile in the middle frame). The profiles in the bottom frame of Fig. 10 indicate a small variation of pressure during the formation of deflagration wave.

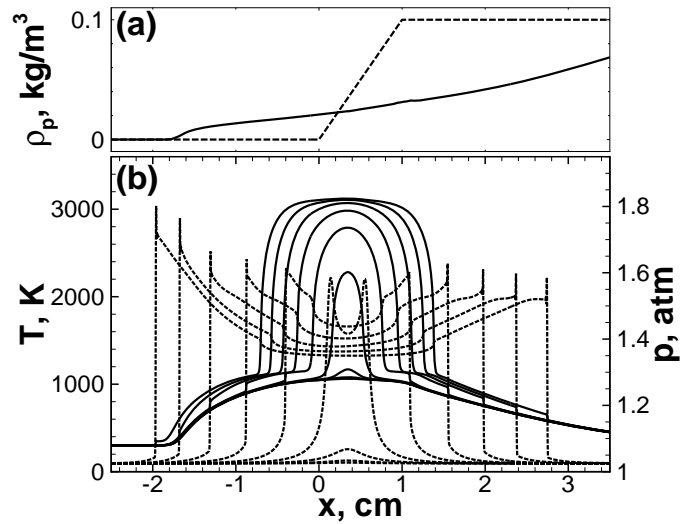

**Figure 11**. Temporal evolution of the gaseous temperature (middle frame), particles mass density (upper frame) and pressure (bottom frame) profiles during the fast combustion wave formation behind the outrunning shock in the vicinity of the margin of the gas-particles cloud ahead of the original propagating flame: $t_0 = 1650 \mu s$, $\Delta \tau = 50 \mu s$. The initial linear density profile of width 1.0cm (dashed line), density profile at $t_0$ prior to the ignition (solid line).

For a particle-gas cloud with a diffuse instead of the stepwise left interface a smoother temperature gradient is formed during the radiation preheating. In the latter case the radiation absorption length varies along the diffusive particle-cloud interface resulting in the formation of the temperature profile with a smoother temperature gradient capable to initiate either fast deflagration or detonation. Examples of the particle-gas cloud with diffuse interface and the computed temperature profiles caused by the radiative preheating are shown in Fig. 11 and Fig. 12. The upper frame in Fig. 11 shows the initial (dashed line) number density distribution of the particles, which drops linearly on the scale 1cm from its maximum value $N_p = 2.5 \cdot 10^7 \text{cm}^{-3}$. The diffuse interface of the particle-cloud is smeared during the radiation preheating due to the expansion of the particle-gas cloud. The temperature gradient with maximum temperature $T^* = 1050\,\text{K}$ is formed at the instant $t_0 = 1600\,\mu\text{s}$ shown by solid line in Fig. 11. The middle frame in Fig. 11 shows the computed time evolution of the gaseous temperature profiles after $t_0 = 1600\,\mu\text{s}$. It depicts the development of spontaneous reaction wave on the formed temperature gradient of the scale $\Delta = (T^* - T_0)/|dT/dx| \approx 8\text{cm}$, which then transits into the fast deflagration behind the outrunning shock in the vicinity of the particle-cloud boundary. The weak shock waves outrunning the deflagration are seen in the bottom frame of Fig. 11.

According to [29, 30] the minimum scale of the initial linear temperature gradient in hydrogen-oxygen mixture at normal conditions ($P_0 = 1\text{atm}$) and $T^* = 1050\,\text{K}$ at the top of the gradient which can initiate a detonation is $(T^* - T_0)/\nabla T \approx 20\text{cm}$. This is possible for the particle-cloud shown in Fig. 12 where on the left diffuse interface of the particles number density drops linearly on the scale of 10cm from its maximum value $N_p = 2.5 \cdot 10^7 \text{cm}^{-3}$. The upper frame in Fig. 12 shows the initial particles number density profile (dashed line) and the formed particles number density due to the expansion during the radiation heating at the instant $t_0 = 4980\,\mu\text{s}$ prior to the ignition time (solid line). Time sequences of the gaseous temperature profiles during the detonation formation in the vicinity of the diffusive interface is depicted in the middle frame in Fig.12.

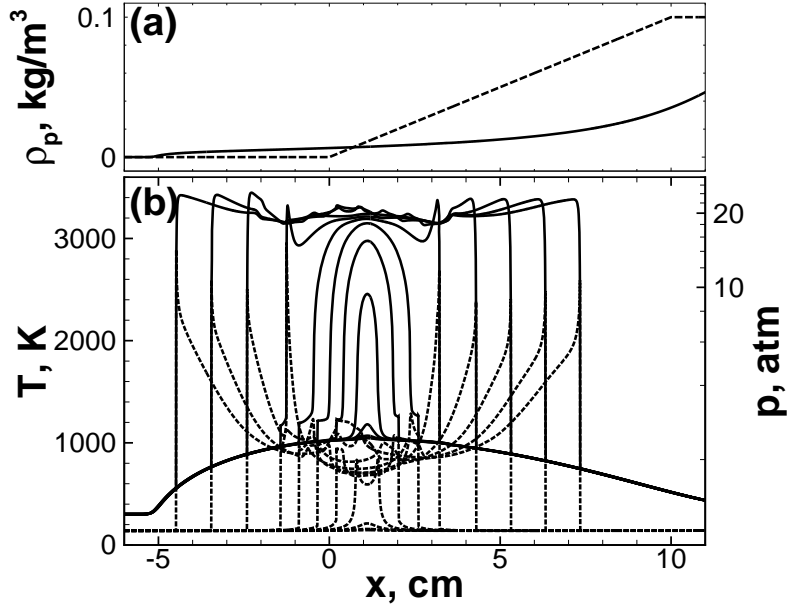

**Figure 12**. Temporal evolution of the gaseous temperature profiles (middle frame) and pressure profiles (bottom frame) during the detonation formation in the vicinity of the gas-particles cloud boundary ahead the propagating flame: $t_0 = 4980\mu s$, $\Delta\tau = 4\mu s$. The upper frame shows the distribution of particles mass density: the initial linear density profile of width 10.0cm (dashed line) and density profile at time instant $t_0 = 4980\mu s$ prior to the ignition (solid line).

One can observe the development of spontaneous reaction wave, which then is coupling with the shock wave and finally due to the positive feedback between the reaction and pressure pulses the detonation wave is formed. Time evolution of the pressure profiles, corresponding to the formation of the shock wave and its coupling with the spontaneous reaction wave, the enhancement and formation of a strong shock corresponding to the transition to detonation is shown in the bottom frame of Fig. 12.

## 6. Conclusions

The present study has been performed in order to examine the role of the thermal radiation emitted from the flame propagating in a particle-gas cloud mixture and to examine the ignition of different combustion modes due to the thermal radiation preheating of the unburned part of the cloud ahead of the original flame. The gaseous phase is assumed to be transparent for the thermal radiation, while the suspended inert particles ahead of the flame absorb and reemit the radiation. The radiant heat is then lost by conduction to the surrounding unreacted gaseous phase so that the gas temperature lags that of the particles. The paper reports a theoretical and DNS studies of the effects of radiation heat transfer on the combustion regimes in particle-laden hydrogen-oxygen and hydrogen-air mixtures. The high resolution direct numerical simulations solve the full system of two phase gas dynamic equations with suspended inert particles with a detailed chemical kinetics for gaseous phase. Two cases have been analyzed: flame propagating in a uniformly dispersed particle-cloud mixture and a flame propagating in non-uniformly dispersed particle-cloud layered mixture. It is shown that depending on the spatial distribution of the dispersed particles and on the value of the radiation absorption length the consequence of the radiative preheating of the mixture ahead of the flame can be either the increase of the flame velocity for the uniformly dispersed particles or ignition a new deflagration or detonation ahead of the original flame via the Zel'dovich gradient mechanism. What kind of combustion regime is ignited in the latter case of a layered particle-cloud deposits depends on the radiation absorption length and correspondingly on the steepness of the formed temperature gradient in the preignition zone that can be treated independently of the primary flame. It has been shown that as soon as combustion has been generated by the primary particle-laden flame, this combustion may generate secondary explosions ahead of the flame in a distant particle-layered deposits. The performed numerical simulations demonstrate the plausibility of radiation preheating as the principal effect of the combustion intensification and in some cases initiation of detonation in the gaseous fuel, where relatively low concentration of suspended particles or any other substance can absorb the radiative heat flux and rise temperature of the fuel ahead of the flame. The obtained results show that the thermal radiative preheating play a significant role in determining the regimes of combustion in two-phase reacting flows, and presumably can explain the origin of dust explosion in processes that are accompanied by the formation of clouds of fine dust particles and represents common risks in the hydrogen and hydrogen-based technologies and other industries.

It should be emphasized that this study is a necessary prerequisite aiming to show principle physics and role of the radiative preheating using a model of plane flame propagating in hydrogen mixtures with suspended particles. The conditions under which a reactive two-phase mixture can ignite and produce a heat release are important in different areas of fire safety. This danger was initially confined to mine coal industries, but nowadays it spread to almost every industry representing common risks in the coal, metallurgy, chemicals, wood, hydrogen and hydrogen-based technologies and other industries. This paper describes the first steps we have taken in developing a multidimensional numerical model to study explosions in large-scale systems containing mixtures of hydrogen and air and extending the study for coal dust as well. This final model must have the ability to compute the different stages of the evolution of a chemically reactive flow: ignition by a small accidental spark, flame acceleration, development of turbulent flow, clustering of particles in turbulent flow [66-69], which can result in non-uniform emissivity of radiation from the flame as well in non-uniform absorption of the radiation, and detonation initiation. Here we describe the first steps toward achieving this objective.

**Acknowledgments**

One of the authors (ML) acknowledges the following colleagues who helped in immeasurable way during writing this paper: Axel Brandenburg, Nils E. Haugen, Igor Rogachevskii, Nathan Kleeorin. Authors acknowledge the allocation of computing resources provided by the Swedish National Allocations Committee at the Center for Parallel Computers at the Royal Institute of Technology in Stockholm, the National Supercomputer Centers in Linkoping and the Nordic Supercomputer Center in Reykjavik. This work was supported in part by Ben-Gurion University Fellowship for senior visiting scientists (ML) and by the Research Council of Norway under the FRINATEK (Grant 231444, ML).

**References**

1. Bentaib, A., Meynet, N., Bleyer, A. (2015). Overview on hydrogen risk research and development activities: Methodology and open issues Nuclear Engineering and Technology. 47, 26–32.

2. Gupta, S. (2015). Experimental investigations relevant for hydrogen and fission product issues raised by the Fukushima accident, Nuclear Engineering and Technology, 47, 11–25.

3. Lees' Loss Prevention in the Process Industries: Hazard Identification, Assessment and Control (2005). Ed. Sam Mannan and Frank P. Lees, Butterworth-Heinemann Publisher.

4. Flame Acceleration and Deflagration-to-Detonation Transition in Nuclear Safety. (2000). State-of-the *Art Report,* OCDE-Nuclear Safety, NEA/CSNI/R.

5. Kikukawa, S., Mitsuhashi, H., Miyake, A. (2009). Risk assessment for liquid hydrogen fueling stations. Int. J. Hydrogen energy, 34, 1135-1141.

6. Pasman, H.J. (2011). Challenges to improve confidence level of risk assessment of hydrogen technologies. Int. J. Hydrogen Energy, 36, 2407–2413.

7. Eichert H, Fischer M. (1986). Combustion-related safety aspects of hydrogen in energy applications. Int. J. Hydrogen Energy, 11, 117-24.

8. Ng, H. D., John H.S. Lee, J.H.S. (2008). Comments on explosion problems for hydrogen safety. J. Loss Prevent. Proc., 21, 136–146.

9. Eckhoff, R. K. (2009). Understanding dust explosions-the role of powder science and technology. J. Loss Prevent. Proc., 22, 105-116.

10. Bleyer, A., Taveau, J., Chaumeix, N., Paillard, C.E., and Bentaib, A. (2010). Experimental and numerical study of flame propagation with hydrogen gradient in a vertical facility: ENACCEF. Presented at the 8th International Symposium on Hazards, Prevention and Mitigation of Industrial Explosions, September 5–10, Yokohama, Japan.

11. Ciccarelli G., Dorofeev S. (2008). Flame acceleration and transition to detonation in ducts, Prog. Energy and Combust. Science, 34, 499-550.

12. Oran, E.S., Gamezo, V.N. (2007). Origins of the deflagration-to-detonation transition in gas-phase combustion, Combustion and Flame, 148, 4–47.

13. Roy, G.D., Frolov, S.M., Borisov, A.A., Netzer, D.W. (2004). Pulse detonation propulsion: challenges, current status, and future perspective. Int. Prog. Energy Combust. Sci. 30, 545-672.

14. Smirnov, N.N., Nikitin, V.F., Boichenko, A.P., Tyurnikov, M.V., Baskakov, V.V. Deflagration-to-detonation transition in gases and its application to pulsed detonation devices. In: Gaseous and heterogeneous detonations, Edt. By Roy GD, Frolov SM, Kailasaath K, and Smirnov N. ENAS Publishers, Moscow 1999.

15. Shchelkin, K.I. (1940). Influence of the wall roughness on initiation and propagation of detonation in gases (in Russian). Zh. Eksp. Teor. Fiz. (in Russian), 10, 823-837.

16. Shchelkin, K. I., Troshin, Ya. K. (1965). Gasdynamics of combustion. Baltimore: Mono Book Corp.

17. Zel'dovich, Ya. B., Kompaneets, A. S. (1960). Theory of detonation. New York: Academic Press.

18. Urtiew, P., Oppenheim, A. K. (1966). Experimental observation of the transition to detonation in an explosive gas. Proc Roy Soc Lond. Ser A, 295, 13-28.

19. Teodorczyk, A., Drobniak, P., Dabkowski, A. (2009). Fast turbulent deflagration and DDT of hydrogen–air mixtures in small obstructed channel. Int. J. Hydrogen Energy, 34, 5887-5893.

20. Wen, X.,Yu, M., Ji, W., Yue, M., Chen, J. (2015). Methane–air explosion characteristics with different obstacle configurations, Int. J. Mining Sci. Techn., 25, in press. Available online.

21. Proust, C. (2015). Gas Flame Acceleration in Long Ducts, J. Loss Prevent. Process Industries, in press, Available online. DOI: 10.1016/j.jlp.2015.04.001 (Tenth International Symposium on Hazards, Prevention, and Mitigation of Industrial Explosions, Bergen, Norway, 10-14 June 2014).

22. Kuznetsov, M., Alekseev, V., Matsukov I., Dorofeev, S. (2005). DDT in a smooth tube filled with a hydrogen-oxygen mixture, Shock Waves, 14, 205-215.

23. Kuznetsov, M., Liberman, M., Matsukov, I. (2010). Experimental study of the preheat zone formation and deflagration-to-detonation transition, Combust. Sci. Techn, 182, 1628-1644.

24. Liberman, M.A., Ivanov, M.F., Kiverin, A.D., Kuznetsov, M. S., Chukalovsky, A.A., Rakhimova, T.V. (2010). Deflagration-to-detonation transition in highly reactive combustible mixtures. Acta Astronautica, 67, 688-701.

25. Wu, M.H., Burke, M.P., Son, S.F., Yetter, R. A. (2007). Flame acceleration and the transition to detonation of stoichiometric ethylene/oxygen in microscale tubes. Proc. Combust. Inst. 31, 2429-2436.

26. Zel'dovich Ya. B. (1947). On the theory on transition to detonation in gases, Zh. Tekhn. Fiz (J. Techn. Phys.) 17, 3-26 (in Russian).

27. Zel'dovich, Ya. B. (1980). Regime classification of an exothermic reaction with nonuniform initial conditions, Combust. Flame, 39, 211-226.

28. Liberman, M.A. (2014). Unsteady Combustion Processes Controlled by Detailed Chemical Kinetics, Notes on Numerical Fluid Mechanics and Multidisciplinary Design, Vol. 127, Active Flow and Combustion Control, 2014, Springer, Editor: Rudibert King ISBN: 978-3-319-11966-3, pp. 317-341.

29. Liberman, M. A., Kiverin, A. D., Ivanov, M. F. (2011). On Detonation Initiation by a Temperature Gradient for a Detailed Chemical Reaction Models, Phys. Letters, A375, 1803-1808.

30. Liberman, M. A., Kiverin, A. D., Ivanov, M. F. (2012). Regimes of chemical reaction waves initiated by nonuniform initial conditions for detailed chemical reaction models, Phys. Rev. E85, 056312/1-9.

31. Ivanov, M.F., Kiverin, A.D., Liberman, M.A. (2011). Hydrogen-oxygen flame acceleration and transition to detonation in channels with no-slip walls for a detailed chemical reaction model, Phys. Rev. E., 83, 056313/1-16.

32. Ivanov, M.F., Kiverin, A.D., Liberman, M.A. (2011). Flame Acceleration and Deflagration-to-Detonation Transition in Stoichiometric Hydrogen/Oxygen in Tubes of Different Diameters, Int. J. Hydrogen Energy. 36, 7714-7728.

33. Ivanov, M.F., Kiverin, A.D., Yakovenko, I.S., Liberman, M.A. (2013). Hydrogen-Oxygen Flame Acceleration and Deflagration-to-Detonation Transition in Three-dimensional Rectangular Channels with no-slip Walls, Int. J. Hydrogen Energy. 38, 16427-16440.

34. Machida, T., Asahara, M., Hayashi, A. K., Tsuboi, N. (2014). Three-Dimensional Simulation of Deflagration-to-Detonation Transition with a Detailed Chemical Reaction Model. Combust. Sci. Technol. 186, 1758–1773.

35. Kiverin, A.D., Kassoy, D.R., Ivanov, M.F., Liberman, M.A. (2013). Mechanisms of Ignition by Transient Energy Deposition: Regimes of Combustion Waves Propagation, Phys. Rev. E87, 033015/1-9.

36. Pekalski, A., Puttock, J., Chynoweth, S. (2015). Deflagration to detonation transition in a vapour cloud explosion in open but congested space: large scale test, J. Loss Prevent. Process Industries. doi: 10.1016/j.jlp.2015.04.002.

37. Essenhigh, R. H., Csaba, J. (1963). The thermal radiation theory for plane flame propagation in coal dust clouds. Symposium (International) on Combustion, 9(1), 111-125.

38. Joulin G., Deshaies B. (1986). On Radiation-Affected Flame Propagation in Gaseous Mixtures Seeded with inert Particles, Combust. Sci. Techn., 47, 299-315.

39. Joulin G., Deshaies B. (1986). Radiative Transfer as a Propagation Mechanism for Rich Flames of Reactive Suspensions, SIAM J. Appl. Math., 46, 561–581.

40. Joulin G., Eudier M. (1989). Radiation-dominated propagation and extinction of slow, particle-laden gaseous flames, Symposium (International) on Combustion, 22, 1579–1585.

41. Choi, S., Kruger, C. H. (1985). Modeling Coal Particle Behavior under Simultaneous Devolatilization and Combustion, Combust. Flame, 6, 131-144.

42. Rockwell S. R., Rangwala A. S. (2013). Influence of coal dust on premixed turbulent methane–air flames, Combust. Flame, 160, 635–640.

43. Beyrau, F., Hadjipanayis, M.A., Lindstedt, R.P. (2013). Ignition of fuel/air mixtures by radiatively heated particles, Proceedings Combust. Institute 34, 2065–2072.

44. Proust, C. (2006). A few fundamental aspects about ignition and flame propagation in dust clouds, J. Loss Prev. Process Ind., 19, 104-120.

45. Cao, W., Gao, W., Liang, J. Xu, S., Pan, F. (2014). Flame-propagation behavior and a dynamic model for the thermal-radiation effects in coal-dust explosions, J. Loss Prev. Process Ind. 29, 65-71.

46. Gao, W., Mogi, T., Yu, J., Yan, X., Sun, J., Dobashi, R. (2015). Flame propagation mechanisms in dust explosions. J. Loss Prevent. Process Ind. in press, http://dx.doi.org/10.1016/j.jlp.2014.12.021.

47. Hadjipanayis, M.A., Beyrau, F., Lindstedt, R.P., Atkinson, G., Cusco, L. (2015). Thermal Radiation from Vapour Cloud Explosions, Process Safety and Environmental Protection, 94,517–527.

48. Bidabadi M., Zadsirjan S., Mostafavi S. A. (2013). Radiation heat transfer in transient dust cloud flame propagation, J. Loss Prevent. Process Ind., 26, 862-868.

49. Dobashi, R., Senda, K. (2006). Detailed analysis of flame propagation during dust explosions by UV band observations, J. Loss Prevent. Process Ind., 19, 149-153.

50. Benedetto, A.D., Agreda, A.G., Dufaud, O., Khalili, I., Sanchirico, R., Cuervo, N., Perrin, L., Russo, P. (2011). Flame propagation of dust and gas-air mixtures in a tube. In: Proceedings of 7th Mediterranean Combustion Symposium, Sardinia, Italy.

51. Warnatz J., Maas U., Dibble R. W. (2001). Combustion. Physical and chemical fundamentals, modeling and simulations, experiments, pollutant formation, Springer.

52. Hirschfelder J. O., Curtiss C. F., Bird R. B. (1964). Molecular theory of gases and liquids, Wiley, New York.

53. Heywood J. B. (1988). Internal combustion engine fundamentals, Mc.Graw-Hill, New York.

54. Acrivos A., Taylor T.D. (1962). Heat and mass transfer from single spheres in Stokes flow, Phys. Fluids. 5, 387-394.

55. Zeldovich Ya. B. and Raizer Yu. P. (1966). Physics of Shock waves and High-Temperature Hydrodynamic Phenonema. Academic Press, New-York-London.

56. Siegel R., Howell J.R. (1993). Thermal Radiation Heat Transfer. (3ed Ed). Taylor &Francis Group, Washington-Philaelphia-London.

57. Gentry, R. A., Martin, R. E., and Daly, B. J. (1966). An Eulerian Differencing Method for Unsteady Compressible Flow Problems, J. Comp. Phys., 1, 87-118.

58. Belotserkovsky O.M., Davydov Yu.M. (1982). Coarse-particle method in hydrodynamics. Russian Publ. Inc. Nauka, Mir, Moscow.

59. Liberman M.A., Eriksson L-E, Ivanov M.F., Piel O.D. (2005). Numerical Modeling of the Propagating Flame and Knock Occurrence in Spark-Ignition Engines. Combust. Sci. Techn. 177, 151-182.

60. Liberman M.A., Ivanov M.F., Valiev D.M. (2006). Numerical modeling of knocking onset and hot spot formation by propagating flame in SI engines. Combust. Sci. Techn. 178, 1613-1647.

61. Hairer E., Wanner G. (1996). Solving ordinary differential equations II. Stiff and differential – algebraic problems. Springer–Verlag, New York.

62. Woodward, P., Colella, P. (1984). The numerical simulation of two-dimensional fluid flow with strong shocks, J. Comp. Phys. 54, 115-173.

63. Zel'dovich, Ya.B., Barenblatt, G. I., Librovich, V. B., Makhviladze, G. M. (1985). The Mathematical Theory of Combustion and Explosion, Consultants Bureau, New York.

64. Bychkov V.V., Liberman M.A. (2000). Dynamics and Stability of Premixed Flames, Physics Reports, vol. 325, 115–237.

65. Bowman, C.T., Hanson, R.K., Lissianski, U., Frenklach, M., Goldenberg, M., Smith, G.P. (1996). GRI-Mech. 2.11—An optimized detailed chemical reaction mechanism for methane combustion and NO formation and reburning, Technical Report, Gas Research Institute, Chicago, IL.

66. Elperin, T., Kleeorin, N., Liberman, M.A., L'vov, V.S., Rogachevskii, I. (2007). Clustering of aerosols in atmospheric turbulent flow. Environ. Fluid Mech. 7, 173-187.

67. Elperin, T., Kleeorin, N., Rogachevskii, I. (1996). Self-excitation of fluctuations of inertial particles concentration in turbulent fluid flow. Phys. Rev. Lett. 77, 5373-5377.

68. Olla, P. (2010). Preferential concentration versus clustering in inertial particle transport by random velocity fields. Phys. Rev. E81, 016305.

69. Elperin, T., Kleeorin, N., Liberman, M.A., Rogachevskii, I. (2013). Tangling clustering instability for small particles in temperature stratified turbulence, Phys. Fluids 25, 085104-1/19 (2013).